\documentclass[cits]{PoS}

\title{Measurements of the Spectral Energy Distribution of the Cosmic Infrared Background}

\ShortTitle{Measurements of the spectral energy distribution of the cosmic infrared background}

\author{\speaker{Matthieu Bethermin}\\
       Institut d'Astrophysique Spatiale, Univ. Paris Sud 11 \& CNRS, Orsay\\
       E-mail: \email{matthieu.bethermin@ias.u-psud.fr}}

\author{Herv\'e Dole\\
        Institut d'Astrophysique Spatiale, Univ. Paris Sud 11 \& CNRS, Orsay, and Institut Universitaire de France\\
        E-mail: \email{herve.dole@ias.u-psud.fr}}

\abstract{The extragalactic background light (EBL) is the relic emission of all processes of structure formation in the Universe. About half of this background, called the Cosmic Infrared Background (CIB) is emitted in the 8-1000 microns range, and peaks around 150 microns. It is due to the dust reemission from star formation processes and AGN emission. The CIB spectral energy distribution (SED) constraints the models of star formation in the Universe. It is also useful to compute the opacity of the Universe to the TeV photons.
We present the different types of measurements of the CIB and discuss their strengths and weaknesses.

1. The absolute SED was measured by COBE, and by other experiments. These measurements are limited by the accuracy of the component separation, i.e. the foreground subtraction.

2. Robust lower limits are determined from the extragalactic number
counts of infrared galaxies. These lower limits are very stringent up to 100 microns. At larger wavelengths, the rather low angular resolution of the instruments limits strongly the depth of the number counts. The "stacking" method determines the flux emitted at a given wavelength by a population detected at another wavelength, and provides stringent lower limits in the sub-mm range. It is complementary with other methods based on the statistical analysis of the map properties like the P(D) analysis.

3. Finally, upper limits can be derived from the high energy spectra of extragalactic sources. These upper limits give currently good constraints in the near- and mid-IR.

Progress have been amazing since the CIB discovery about 15 years ago: the SED is much better known, and most of it can be accounted for by galaxies (directly or indirectly). Prospects are also exciting, with fluctuation analysis with Planck \& Herschel, and forthcoming missions.}

\FullConference{Cosmic Radiation Fields: Sources in the early Universe - CRF2010,\\
		November 9-12, 2010\\
		Desy Germany}

\begin{document}

\section{Introduction}

The cosmic infrared background (CIB) was detected for the first time in the late nineties in the FIRAS and DIRBE data \cite{Puget1996,Fixsen1998,Hauser1998}. This background, lying in the mid- far-infrared and submillimeter range (usually defined between 8 and 1000~$\mu$m), represents about half \cite{Dole2006} of the extragalactic background light (EBL), which is the relic of all the structure formation processes. The CIB emission is mainly due to the outputs of the infrared galaxies, and in a minor way to the obscured active galactic nuclei (AGN) \cite{Hauser2001,Lagache2005}. It peaks near 150~$\mu$m .\\

The output of the infrared galaxies is a good probe of the star formation \cite{Kennicutt1998}. During a starburst, the massive stars have short lives but strong UV emissions, which are absorbed and reprocessed by the dusty environment where they formed. Infrared outputs are thus a probe of the presence of massive young stars and thus of the recent star formation in a galaxy. Dust heated by AGN also contribute to the CIB, but almost an order of magnitude fainter than starbursts \cite{Jauzac2010}. The spectral energy distribution (SED) of the CIB thus constrains the star formation history and the galaxy evolution models \cite{LeBorgne2009,Valiante2009,Franceschini2010,Bethermin2010c,Marsden2010}.

\section{Direct measurements}

Direct measurements of the CIB can be performed using absolute photometry, but these measurements are affected by the accuracy of the foreground modeling. These foregrounds are the zodiacal emission \cite{Kelsall1998} and the galactic cirrus \cite{Lagache2000}. The zodiacal emission is the thermal radiation of the interplanetary dust. The galactic cirrus are diffuse clouds of dust in our galaxy heated by the UV emissions of the stars. At 20~$\mu$m, the zodiacal light is three order of magnitude brighter than the CIB. The accuracy of the zodiacal subtraction is thus the main limitation of the absolute measurements. The galactic cirrus output have the same order of magnitude than the CIB. They can be removed accurately using the current HI data \cite{Penin2010}. At larger wavelength, the cosmic microwave background have to be subtracted. The spectrum of the CMB being well known, its subtraction can be done accurately \cite{Fixsen2009}.

\begin{figure}
\centering
\includegraphics{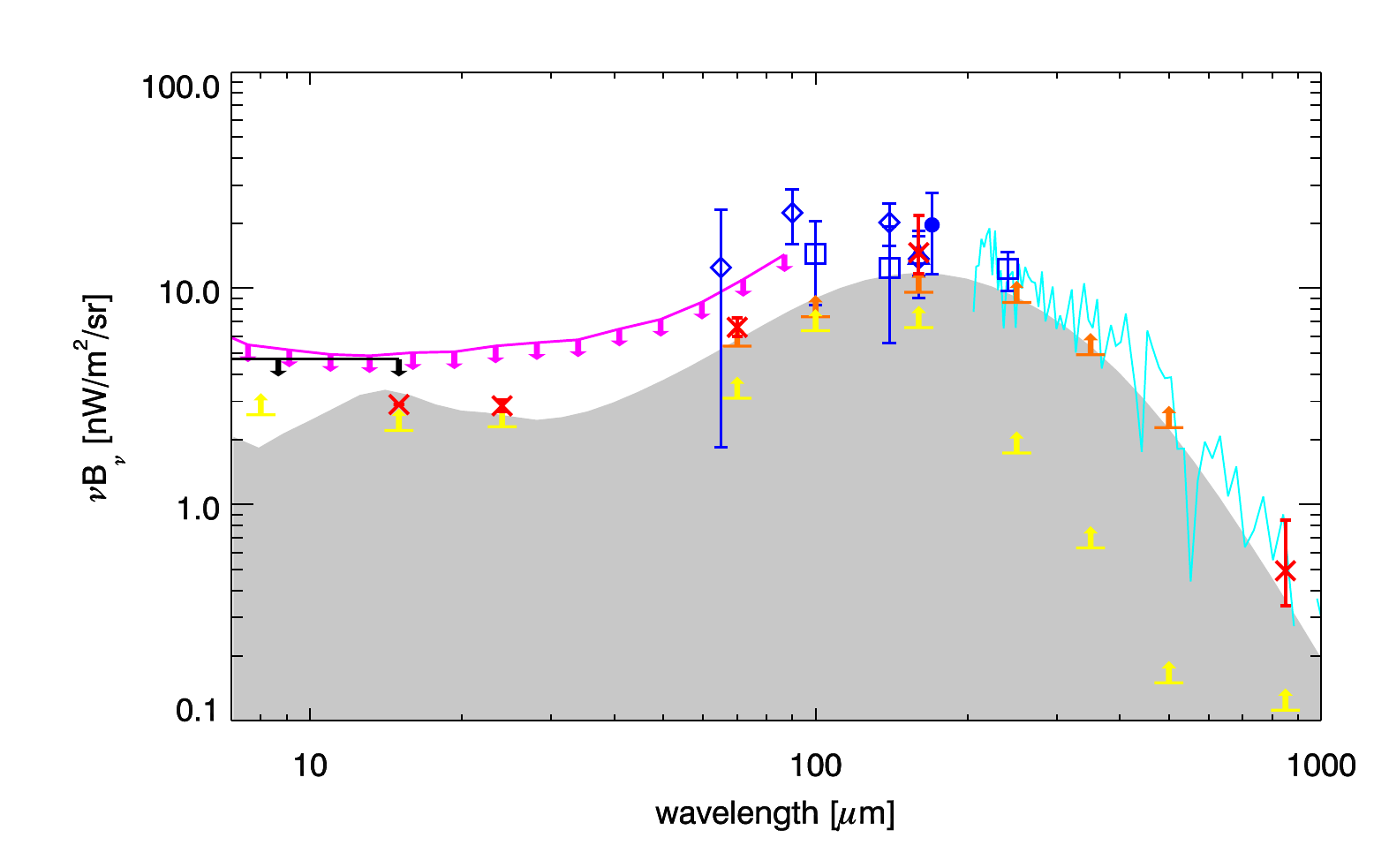}
\caption{\label{fig:cib} Spectral energy distribution of the cosmic infrared background. \textit{Yellow arrows}: Lower limits from resolved source counts at 8~$\mu$m \cite{Fazio2004}, 15~$\mu$m \cite{Teplitz2010}, 24~$\mu$m, 70~$\mu$m \cite{Bethermin2010a}, 100~$\mu$m, 160~$\mu$m \cite{Berta2010}, 250~$\mu$m, 350~$\mu$m and 500~$\mu$m \cite{Oliver2010} and 850~$\mu$m \cite{Coppin2008}. \textit{Orange arrows}: Lower limits from stacking analysis at 70~$\mu$m \cite{Bethermin2010a}, 100~$\mu$m, 160~$\mu$m \cite{Berta2010}, 250~$\mu$m, 350~$\mu$m and 500~$\mu$m \cite{Marsden2009}. \textit{Red crosses}: Extrapolated total contribution of the galaxies to the CIB at 15~$\mu$m \cite{Teplitz2010}, 24~$\mu$m, 70~$\mu$m, 160~$\mu$m \cite{Bethermin2010a} and 850~$\mu$m \cite{Zemcov2010}. \textit{Blue square}: DIRBE absolute measurements \cite{Hauser1998}. \textit{Blue diamonds}: Akari absolute measurements \cite{Matsuura2009}. \textit{Blue triangle}: MIPS absolute measurement at 160~$\mu$m \cite{Penin2010}. \textit{Blue circle}: ISOPHOT absolute measurement at 170~$\mu$m \cite{Juvela2009}. \textit{Cyan solid line}: FIRAS Spectrum \cite{Lagache2000}. \textit{Black arrows}: Upper limits from TeV opacity of the Universe \cite{Renault2001}. \textit{Purple arrows}: Upper limits from TeV opacity of the Universe \cite{Mazin2007}. \textit{Light grey area}: CIB predicted by the Bethermin et al. model \cite{Bethermin2010c}.}
\end{figure}

First measurements of the CIB was performed with DIRBE \cite{Hauser1998} and FIRAS \cite{Fixsen1998,Lagache2000} onboard COBE. The DIRBE measurements have been updated in \cite{Dole2006} (see discussion about DIRBE/FIRAS cross-calibration and subtraction of the zodiacal light). More recent measurements have been performed with, ISO \cite{Juvela2009}, Akari \cite{Matsuura2009} and \textit{Spitzer} \cite{Penin2010}. These measurements are summarized in the Fig. \ref{fig:cib} and Table \ref{tab:cib}. Notice the large scatter in the measurements. Notice also the IRAS measurements \cite{Miville2002} based on a fluctuation analysis.

\section{Lower limits from the number counts}

In the mid-infrared, the depth and the angular resolution of the recent infrared surveys is sufficient to resolve the main part of the CIB into sources. Stringent lower limits can thus be derived from the integration of the source counts. These lower limits were derived at 8~$\mu$m using the \textit{Spitzer}/IRAC data \cite{Fazio2004}, at 15~$\mu$m using ISO \cite{Elbaz1999}, \textit{Spitzer} \cite{Teplitz2010} and Akari \cite{Pearson2010} data, and at 24~$\mu$m \cite{Bethermin2010a}. Estimations of the total contribution of the galaxies to the CIB was done by extrapolating the faint-end slope of the counts. Assuming the CIB is only due to the galaxies, the fraction of the CIB resolved directly is then estimated to be $\sim$80\% \cite{Bethermin2010a}.\\

At larger wavelengths, the angular resolution of the instruments decreases compared to wavelength, and it is harder to resolve a significant part of the CIB due to the confusion \cite{Dole2003,Dole2004}. For example, we resolve only 40\% and 15\% of the CIB at 70 and 160~$\mu$m, respectively with \textit{Spitzer} \cite{Bethermin2010a}. Thanks to its 3.5~m diameter mirror, \textit{Herschel} resolves $\sim$45\% and $\sim$52\% at 100 and 160~$\mu$m, respectively \cite{Berta2010}. Nevertheless, this fraction decreases to $\sim$15\%, $\sim$10\% and $\sim$6\% at 250, 350 and 500~$\mu$m, respectively \cite{Oliver2010}. At 850~$\mu$m, the opacity of the atmosphere is sufficiently low to allow observations from the ground. The counts at this wavelength resolve only 20-30\% of the CIB \cite{Coppin2008}. Deeper counts, resolving the bulk of the CIB, are derived in fields lensed by low-z galaxy clusters, where the error budget is dominated by the large scale structure (narrow field) \cite{Altieri2010,Zemcov2010}. These measurements are summarized in the Fig.~\ref{fig:cib} and Table~\ref{tab:cib}.

\begin{table}
\centering
\begin{tabular}{lllll}
\hline
\hline
Wavelength & Reference & CIB level & Instrument & Comments \\
$\mu$m & & nW.m$^{-2}$.sr$^{-1}$ & & \\
\hline
\multicolumn{5}{c}{Absolute measurements}\\
\hline

65 & Matsuura et al. (2009) & 12.5$\pm$1.4$\pm$9.2 & Akari & \\
90 & Matsuura et al. (2009) & 22.3$\pm$1.7$\pm$4.7 & Akari & \\
100 & Lagache et al. (2000)& 14.4$\pm$6.0 & DIRBE & Updated in Dole et al. (2006)\\
140 &  Lagache et al. (2000)& 12.4$\pm$6.9 & DIRBE & Updated in Dole et al. (2006)\\
140 & Matsuura et al. (2009) & 20.1$\pm$3.4$\pm$1.1 & Akari & \\
160 & Matsuura et al. (2009) & 13.7$\pm$3.9$\pm$40.8 & Akari & \\
160 & Penin et al. (2011) & 14.43$\pm$3 & \textit{Spitzer}/MIPS & \\
170 & Juvela et al. (2009) & 19.6$\pm$5.8$\pm$5.5 & ISOPHOT & \\
240 & Lagache et al. (2000) & 12.3$\pm$2.5 & DIRBE & Updated in Dole et al. (2006)\\
200-1000 &  Lagache et al. (2000) & see figure & FIRAS  & \\
\hline
\multicolumn{5}{c}{Lower limit from resolved sources}\\
\hline
8 & Fazio et al. (2004) & 5.4 & \textit{Spitzer}/IRAC & QSO1700+EGS+Bootes\\
15 & Teplitz et al. (2010) & 2.2$\pm$0.2 & \textit{Spitzer}/IRS & GOODS-N and S\\
24 & B\'ethermin et al. (2010) &  2.29$\pm$0.09 & \textit{Spitzer}/MIPS & FIDEL+COSMOS+SWIRE\\
70 & B\'ethermin et al. (2010) &  3.1$\pm$0.2 & \textit{Spitzer}/MIPS & FIDEL+COSMOS+SWIRE\\
100 & Berta et al. (2010) & 6.33$\pm$1.67 & \textit{Herschel}/PACS & SDP data\\
160 & Berta et al. (2010) & 6.58$\pm$1.62 & \textit{Herschel}/PACS & SDP data\\
250 & Oliver et al. (2010) & 1.73$\pm$0.33 & \textit{Herschel}/SPIRE & SDP data\\
350 & Oliver et al. (2010) & 0.63$\pm$0.18 & \textit{Herschel}/SPIRE & SDP data\\
500 & Oliver et al. (2010) & 0.15$\pm$0.07 & \textit{Herschel}/SPIRE & SDP data\\
850 & Coppin et al. (2008) & 0.11$_{-0.04}^{+0.05}$ & SCUBA & SHADES survey\\
\hline
\multicolumn{5}{c}{Lower limits from stacking analysis}\\
\hline
70 & B\'ethermin et al. (2010) & 5.4$\pm$0.4 & \textit{Spitzer}/MIPS & FIDEL+COSMOS+SWIRE\\
100 & Berta et al. (2010) & 7.4$\pm$0.5  & \textit{Herschel}/PACS & SDP data\\
160 & Berta et al. (2010) & 9.6$\pm$0.7 & \textit{Herschel}/PACS & SDP data\\
250 & Marsden et al. (2009) & 8.6$\pm$0.6 & BLAST & biased by clustering ($\sim$+9\%) \\ 
350 & Marsden et al. (2009) & 4.9$\pm$0.3 & BLAST & biased by clustering ($\sim$+13\%) \\ 
500 & Marsden et al. (2009) & 2.3$\pm$0.2 & BLAST & biased by clustering ($\sim$+24\%) \\ 
\hline
\multicolumn{5}{c}{Extrapolated total contribution of the galaxies to the CIB}\\
24 & B\'ethermin et al. (2010) & 2.86$_{-0.16}^{+0.19}$ & \textit{Spitzer}/MIPS & direct extraction + stacking\\
70 & B\'ethermin et al. (2010) & 6.6$_{-0.6}^{+0.7}$ & \textit{Spitzer}/MIPS & direct extraction + stacking\\
160 & B\'ethermin et al. (2010) & 14.6$_{-2.9}^{+7.1}$ & \textit{Spitzer}/MIPS & direct extraction + stacking\\
850 & Zemcov et al. (2010) &  0.34-0.85 & SCUBA & lensing survey\\
\hline
\multicolumn{5}{c}{Upper limits from $\gamma$-rays absorption}\\
\hline
5-15 & Renault et al. (2001) & 4.7 & CAT \& HEGRA & 0.5-20~TeV\\
1-90 & Mazin\&Raue (2007) & 5-40 & TeV exp.& 0.1-20~TeV\\
\hline
\end{tabular}
\caption{\label{tab:cib} Summary of the measurements of the cosmic infrared background.}
\end{table}

\section{More stringent lower limits from the statistical analysis of far-infrared maps}

In the 200-800~$\mu$m range, the faint sources which contribute to the CIB cannot be resolved due to the limited angular resolution of current instruments. Nevertheless, most of these sources are detected in the deepest \textit{Spitzer} 24~$\mu$m surveys. The collective signal of these sources, not detected individually in the FIR and sub-mm, can be detected collectively by stacking, providing very stringent lower limits on the CIB in \textit{Spitzer}/MIPS \cite{Dole2006,Bethermin2010a}, BLAST \cite{Marsden2009}, PACS \cite{Berta2010} bands. It is also possible to compute deep number counts from stacking analysis and to derive an estimation of the total value of the CIB extrapolating the faint end-slope of the number counts \cite{Bethermin2010a}.\\

Ultra-deep counts can also be derived from the analysis of the pixel histograms of the infrared maps --  P(D) analysis \cite{Condon1974,Friedmann2004}. If the clustering of the galaxies is neglected and the sources are point-like, the pixel histogram depends only on the counts and instrumental properties (instrumental noise and PSF). A non-physical broken power-law description of the number counts can thus be fitted to this pixel histogram \cite{Patanchon2009,Glenn2010}. This method provide deep counts which explains about 2/3 of the CIB at 250, 350 and 500~$\mu$m, and is complementary to the stacking analysis. These measurements are summarized in the Fig. \ref{fig:cib} and Table \ref{tab:cib}.

\section{Upper limits from the gamma-ray spectra}

Two photons can interact if the energy in the barycentric frame is sufficient to produce a positron-electron pair. The high-energy $\gamma$ photons can thus be absorbed by the photons of the EBL \cite{Fazio1970,Raue2009}. This absorption can be detected as a break at high energy in the spectrum of high-redshift blazars. This phenomenon provides upper limits on the COB and the CIB \cite{Renault2001}. Using the spectrum of several blazars, it is now possible to derive stringent upper limits to the CIB below 100~$\mu$m \cite{Mazin2007}. The method is somehow dependant on the assumed blazar spectra and on the evolution of the CIB with the redshift \cite{Kneiske2010}, but is extremely powerful in particular at near-infrared wavelengths (not covered here).\\

\section{Conclusion}

The direct measurements of the CIB are difficult because of the accuracy requirements on the foreground subtraction, despite a better knowledge and refined techniques. The lower limits coming from the deep surveys and the upper limits coming from the blazars spectra are more and more stringent, and the association of these two measurements challenges the absolute measurements, especially in the mid-IR, where the zodiacal emission is very strong. Other constraints, e.g. involving the build-up of the CIB with redshift \cite{Jauzac2010} will allow even better observational constraints on the CIB SED and its history, not mentioning the detailed study of the CIB fluctuations (e.g. with Planck at large angular scales \cite{Planck2011} and Herschel at small scales \cite{Amblard2011}, following earlier works e.g. with IRAS, ISO, Spitzer, BLAST, Akari). The progress has been amazing since the CIB first detection 15 years ago, the SED being much better known (despite large uncertainties at some wavelengths, like between 30 and 120 microns), and most of the emission being explained by directly or indirectly detected galaxy populations, mainly luminous and ultraluminous infrared galaxies.\\

In the future, these observational constraints will be significantly improved thanks to the next generation of telescopes. In the mid-IR and far-IR, JWST (6.5~m space telescope) will be able to resolve directly almost all the CIB at those wavelengths. These data will also be useful to perform deeper stacking analysis in the sub-mm. Finally, CCAT (25~m sub-mm ground-telescope) and ALMA (sub-mm interferometer) will be able to resolve directly the bulk of the CIB in the sub-mm domain \cite{Bethermin2010c}, while awaiting possible future space missions dedicated to the study of CIB fluctuations and/or galaxies (e.g. space interferometers in the FIR).

\bibliographystyle{JHEP}
\bibliography{bethermin_CRF2010}

\providecommand{\href}[2]{#2}\begingroup\raggedright\begin{thebibliography}{10}

\bibitem{Puget1996}
J.~{Puget}, A.~{Abergel}, J.~{Bernard}, F.~{Boulanger}, W.~B. {Burton},
  F.~{Desert}, and D.~{Hartmann}, {\it {Tentative detection of a cosmic
  far-infrared background with COBE.}},  {\em \aap} {\bf 308} (Apr., 1996) L5+.

\bibitem{Fixsen1998}
D.~J. {Fixsen}, E.~{Dwek}, J.~C. {Mather}, C.~L. {Bennett}, and R.~A. {Shafer},
  {\it {The Spectrum of the Extragalactic Far-Infrared Background from the COBE
  FIRAS Observations}},  {\em \apj} {\bf 508} (Nov., 1998) 123--128,
  [\href{http://xxx.lanl.gov/abs/astro-ph/}{{\tt astro-ph/}}].

\bibitem{Hauser1998}
M.~G. {Hauser}, R.~G. {Arendt}, T.~{Kelsall}, E.~{Dwek}, N.~{Odegard}, J.~L.
  {Weiland}, H.~T. {Freudenreich}, W.~T. {Reach}, R.~F. {Silverberg}, S.~H.
  {Moseley}, Y.~C. {Pei}, P.~{Lubin}, J.~C. {Mather}, R.~A. {Shafer}, G.~F.
  {Smoot}, R.~{Weiss}, D.~T. {Wilkinson}, and E.~L. {Wright}, {\it {The COBE
  Diffuse Infrared Background Experiment Search for the Cosmic Infrared
  Background. I. Limits and Detections}},  {\em \apj} {\bf 508} (Nov., 1998)
  25--43, [\href{http://xxx.lanl.gov/abs/astro-ph/}{{\tt astro-ph/}}].

\bibitem{Dole2006}
H.~{Dole}, G.~{Lagache}, J.~{Puget}, K.~I. {Caputi}, N.~{Fern{\'a}ndez-Conde},
  E.~{Le Floc'h}, C.~{Papovich}, P.~G. {P{\'e}rez-Gonz{\'a}lez}, G.~H. {Rieke},
  and M.~{Blaylock}, {\it {The cosmic infrared background resolved by Spitzer.
  Contributions of mid-infrared galaxies to the far-infrared background}},
  {\em \aap} {\bf 451} (May, 2006) 417--429,
  [\href{http://xxx.lanl.gov/abs/astro-ph/}{{\tt astro-ph/}}].

\bibitem{Hauser2001}
M.~G. {Hauser} and E.~{Dwek}, {\it {The Cosmic Infrared Background:
  Measurements and Implications}},  {\em \araa} {\bf 39} (2001) 249--307,
  [\href{http://xxx.lanl.gov/abs/astro-ph/}{{\tt astro-ph/}}].

\bibitem{Lagache2005}
G.~{Lagache}, J.~{Puget}, and H.~{Dole}, {\it {Dusty Infrared Galaxies: Sources
  of the Cosmic Infrared Background}},  {\em \araa} {\bf 43} (Sept., 2005)
  727--768, [\href{http://xxx.lanl.gov/abs/astro-ph/}{{\tt astro-ph/}}].

\bibitem{Kennicutt1998}
R.~C. {Kennicutt}, Jr., {\it {The Global Schmidt Law in Star-forming
  Galaxies}},  {\em \apj} {\bf 498} (May, 1998) 541--+,
  [\href{http://xxx.lanl.gov/abs/astro-ph/}{{\tt astro-ph/}}].

\bibitem{Jauzac2010}
M.~{Jauzac}, H.~{Dole}, E.~{Le Floc'h}, H.~{Aussel}, K.~{Caputi}, O.~{Ilbert},
  M.~{Salvato}, N.~{Bavouzet}, A.~{Beelen}, M.~{B{\'e}thermin}, J.~{Kneib},
  G.~{Lagache}, and J.~{Puget}, {\it {The cosmic far-infrared background
  buildup since redshift 2 at 70 and 160 microns in the COSMOS and GOODS
  fields}},  {\em \aap} {\bf 525} (Jan., 2011) A52+,
  [\href{http://xxx.lanl.gov/abs/1009.0419}{{\tt arXiv:1009.0419}}].

\bibitem{LeBorgne2009}
D.~{Le Borgne}, D.~{Elbaz}, P.~{Ocvirk}, and C.~{Pichon}, {\it {Cosmic
  star-formation history from a non-parametric inversion of infrared galaxy
  counts}},  {\em \aap} {\bf 504} (Sept., 2009) 727--740,
  [\href{http://xxx.lanl.gov/abs/0901.3783}{{\tt arXiv:0901.3783}}].

\bibitem{Valiante2009}
E.~{Valiante}, D.~{Lutz}, E.~{Sturm}, R.~{Genzel}, and E.~L. {Chapin}, {\it {A
  Backward Evolution Model for Infrared Surveys: The Role of AGN- and Color-$L_{TIR}$ Distributions}},  {\em \apj} {\bf 701} (Aug., 2009) 1814--1838,
  [\href{http://xxx.lanl.gov/abs/0906.4110}{{\tt arXiv:0906.4110}}].

\bibitem{Franceschini2010}
A.~{Franceschini}, G.~{Rodighiero}, M.~{Vaccari}, S.~{Berta}, L.~{Marchetti},
  and G.~{Mainetti}, {\it {Galaxy evolution from deep multi-wavelength infrared
  surveys: a prelude to Herschel}},  {\em \aap} {\bf 517} (July, 2010) A74+.

\bibitem{Bethermin2010c}
M.~{B{\'e}thermin}, H.~{Dole}, G.~{Lagache}, D.~{Le Borgne}, and
  A.~{P{\'e}nin}, {\it {Modeling the evolution of infrared galaxies: A
  Parametric backwards evolution model}},  {\em ArXiv e-prints} (Oct., 2010)
  [\href{http://xxx.lanl.gov/abs/1010.1150}{{\tt arXiv:1010.1150}}].

\bibitem{Marsden2010}
G.~{Marsden}, E.~L. {Chapin}, M.~{Halpern}, G.~{Patanchon}, D.~{Scott},
  M.~D.~P. {Truch}, E.~{Valiante}, M.~P. {Viero}, and D.~V. {Wiebe}, {\it {A
  Monte Carlo Approach to Evolution of the Far-Infrared Luminosity Function
  with BLAST}},  {\em ArXiv e-prints} (Oct., 2010)
  [\href{http://xxx.lanl.gov/abs/1010.1176}{{\tt arXiv:1010.1176}}].

\bibitem{Kelsall1998}
T.~{Kelsall}, J.~L. {Weiland}, B.~A. {Franz}, W.~T. {Reach}, R.~G. {Arendt},
  E.~{Dwek}, H.~T. {Freudenreich}, M.~G. {Hauser}, S.~H. {Moseley}, N.~P.
  {Odegard}, R.~F. {Silverberg}, and E.~L. {Wright}, {\it {The COBE Diffuse
  Infrared Background Experiment Search for the Cosmic Infrared Background. II.
  Model of the Interplanetary Dust Cloud}},  {\em \apj} {\bf 508} (Nov., 1998)
  44--73, [\href{http://xxx.lanl.gov/abs/astro-ph/}{{\tt astro-ph/}}].

\bibitem{Lagache2000}
G.~{Lagache}, L.~M. {Haffner}, R.~J. {Reynolds}, and S.~L. {Tufte}, {\it
  {Evidence for dust emission in the Warm Ionised Medium sing WHAM data}},
  {\em \aap} {\bf 354} (Feb., 2000) 247--252,
  [\href{http://xxx.lanl.gov/abs/astro-ph/}{{\tt astro-ph/}}].

\bibitem{Penin2010}
A.~{Penin} and {al.}, {\it {An accurate cirrus-free measurement of the
  intensity and anisotropies of the Cosmic Infrared Background at 100 and 160
  microns}},  {\em submitted to A\&A} (2010).

\bibitem{Fixsen2009}
D.~J. {Fixsen}, {\it {The Temperature of the Cosmic Microwave Background}},
  {\em \apj} {\bf 707} (Dec., 2009) 916--920,
  [\href{http://xxx.lanl.gov/abs/0911.1955}{{\tt arXiv:0911.1955}}].

\bibitem{Juvela2009}
M.~{Juvela}, K.~{Mattila}, D.~{Lemke}, U.~{Klaas}, C.~{Leinert}, and C.~{Kiss},
  {\it {Determination of the cosmic far-infrared background level with the
  ISOPHOT instrument}},  {\em \aap} {\bf 500} (June, 2009) 763--768,
  [\href{http://xxx.lanl.gov/abs/0904.2997}{{\tt arXiv:0904.2997}}].

\bibitem{Matsuura2009}
S.~{Matsuura}, M.~{Shirahata}, M.~{Kawada}, T.~T. {Takeuchi}, D.~{Burgarella},
  D.~L. {Clements}, W.~{Jeong}, H.~{Hanami}, S.~A. {Khan}, H.~{Matsuhara},
  T.~{Nakagawa}, S.~{Oyabu}, C.~P. {Pearson}, A.~{Pollo}, S.~{Serjeant},
  T.~{Takagi}, and G.~{White}, {\it {Detection of the Cosmic Far-Infrared
  Background in the AKARI Deep Field South}},  {\em ArXiv e-prints} (Feb.,
  2010) [\href{http://xxx.lanl.gov/abs/1002.3674}{{\tt arXiv:1002.3674}}].

\bibitem{Miville2002}
M.~{Miville-Desch{\^e}nes}, G.~{Lagache}, and J.~{Puget}, {\it {Power spectrum
  of the cosmic infrared background at 60 and 100 $\backslash$umwith IRAS}},
  {\em \aap} {\bf 393} (Oct., 2002) 749--756,
  [\href{http://xxx.lanl.gov/abs/astro-ph/}{{\tt astro-ph/}}].

\bibitem{Fazio2004}
G.~G. {Fazio}, M.~L.~N. {Ashby}, P.~{Barmby}, J.~L. {Hora}, J.~{Huang}, M.~A.
  {Pahre}, Z.~{Wang}, S.~P. {Willner}, R.~G. {Arendt}, S.~H. {Moseley},
  M.~{Brodwin}, P.~{Eisenhardt}, D.~{Stern}, E.~V. {Tollestrup}, and E.~L.
  {Wright}, {\it {Number Counts at 3 $\mu$m $<\lambda<$ 10 $\mu$m from the Spitzer Space Telescope}},  {\em \apjs} {\bf 154} (Sept.,
  2004) 39--43, [\href{http://xxx.lanl.gov/abs/astro-ph/}{{\tt astro-ph/}}].

\bibitem{Teplitz2010}
H.~I. {Teplitz}, R.~{Chary}, D.~{Elbaz}, M.~{Dickinson}, C.~{Bridge},
  J.~{Colbert}, E.~{Le Floc'h}, D.~T. {Frayer}, J.~H. {Howell}, D.~C. {Koo},
  C.~{Papovich}, A.~{Phillips}, C.~{Scarlata}, B.~{Siana}, H.~{Spinrad}, and
  D.~{Stern}, {\it {Spitzer IRS 16 micron Observations of the GOODS Fields}},
  {\em ArXiv e-prints} (Oct., 2010)
  [\href{http://xxx.lanl.gov/abs/1010.1797}{{\tt arXiv:1010.1797}}].

\bibitem{Bethermin2010a}
M.~{B{\'e}thermin}, H.~{Dole}, A.~{Beelen}, and H.~{Aussel}, {\it {Spitzer deep
  and wide legacy mid- and far-infrared number counts and lower limits of
  cosmic infrared background}},  {\em \aap} {\bf 512} (Mar., 2010) A78+,
  [\href{http://xxx.lanl.gov/abs/1001.0896}{{\tt arXiv:1001.0896}}].

\bibitem{Berta2010}
S.~{Berta}, B.~{Magnelli}, D.~{Lutz}, B.~{Altieri}, H.~{Aussel}, P.~{Andreani},
  O.~{Bauer}, A.~{Bongiovanni}, A.~{Cava}, J.~{Cepa}, A.~{Cimatti}, E.~{Daddi},
  H.~{Dominguez}, D.~{Elbaz}, H.~{Feuchtgruber}, N.~M. {F{\"o}rster Schreiber},
  R.~{Genzel}, C.~{Gruppioni}, R.~{Katterloher}, G.~{Magdis}, R.~{Maiolino},
  R.~{Nordon}, A.~M. {P{\'e}rez Garc{\'{\i}}a}, A.~{Poglitsch}, P.~{Popesso},
  F.~{Pozzi}, L.~{Riguccini}, G.~{Rodighiero}, A.~{Saintonge}, P.~{Santini},
  M.~{Sanchez-Portal}, L.~{Shao}, E.~{Sturm}, L.~J. {Tacconi}, I.~{Valtchanov},
  M.~{Wetzstein}, and E.~{Wieprecht}, {\it {Dissecting the cosmic infra-red
  background with Herschel/PEP}},  {\em \aap} {\bf 518} (July, 2010) L30+,
  [\href{http://xxx.lanl.gov/abs/1005.1073}{{\tt arXiv:1005.1073}}].

\bibitem{Oliver2010}
S.~J. {Oliver}, L.~{Wang}, A.~J. {Smith}, B.~{Altieri}, A.~{Amblard},
  V.~{Arumugam}, R.~{Auld}, H.~{Aussel}, T.~{Babbedge}, A.~{Blain}, J.~{Bock},
  A.~{Boselli}, V.~{Buat}, D.~{Burgarella}, N.~{Castro-Rodr{\'{\i}}guez},
  A.~{Cava}, P.~{Chanial}, D.~L. {Clements}, A.~{Conley}, L.~{Conversi},
  A.~{Cooray}, C.~D. {Dowell}, E.~{Dwek}, S.~{Eales}, D.~{Elbaz}, M.~{Fox},
  A.~{Franceschini}, W.~{Gear}, J.~{Glenn}, M.~{Griffin}, M.~{Halpern},
  E.~{Hatziminaoglou}, E.~{Ibar}, K.~{Isaak}, R.~J. {Ivison}, G.~{Lagache},
  L.~{Levenson}, N.~{Lu}, S.~{Madden}, B.~{Maffei}, G.~{Mainetti},
  L.~{Marchetti}, K.~{Mitchell-Wynne}, A.~M.~J. {Mortier}, H.~T. {Nguyen},
  B.~{O'Halloran}, A.~{Omont}, M.~J. {Page}, P.~{Panuzzo}, A.~{Papageorgiou},
  C.~P. {Pearson}, I.~{P{\'e}rez-Fournon}, M.~{Pohlen}, J.~I. {Rawlings},
  G.~{Raymond}, D.~{Rigopoulou}, D.~{Rizzo}, I.~G. {Roseboom},
  M.~{Rowan-Robinson}, M.~{S{\'a}nchez Portal}, R.~{Savage}, B.~{Schulz},
  D.~{Scott}, N.~{Seymour}, D.~L. {Shupe}, J.~A. {Stevens}, M.~{Symeonidis},
  M.~{Trichas}, K.~E. {Tugwell}, M.~{Vaccari}, E.~{Valiante}, I.~{Valtchanov},
  J.~D. {Vieira}, L.~{Vigroux}, R.~{Ward}, G.~{Wright}, C.~K. {Xu}, and
  M.~{Zemcov}, {\it {HerMES: SPIRE galaxy number counts at 250, 350, and 500
  {$\mu$}m}},  {\em \aap} {\bf 518} (July, 2010) L21+,
  [\href{http://xxx.lanl.gov/abs/1005.2184}{{\tt arXiv:1005.2184}}].

\bibitem{Coppin2008}
K.~{Coppin}, M.~{Halpern}, D.~{Scott}, C.~{Borys}, J.~{Dunlop}, L.~{Dunne},
  R.~{Ivison}, J.~{Wagg}, I.~{Aretxaga}, E.~{Battistelli}, A.~{Benson},
  A.~{Blain}, S.~{Chapman}, D.~{Clements}, S.~{Dye}, D.~{Farrah}, D.~{Hughes},
  T.~{Jenness}, E.~{van Kampen}, C.~{Lacey}, A.~{Mortier}, A.~{Pope},
  R.~{Priddey}, S.~{Serjeant}, I.~{Smail}, J.~{Stevens}, and M.~{Vaccari}, {\it
  {The SCUBA HAlf Degree Extragalactic Survey - VI. 350-{$\mu$}m mapping of
  submillimetre galaxies}},  {\em \mnras} {\bf 384} (Mar., 2008) 1597--1610,
  [\href{http://xxx.lanl.gov/abs/0711.0274}{{\tt arXiv:0711.0274}}].

\bibitem{Marsden2009}
G.~{Marsden}, P.~A.~R. {Ade}, J.~J. {Bock}, E.~L. {Chapin}, M.~J. {Devlin},
  S.~R. {Dicker}, M.~{Griffin}, J.~O. {Gundersen}, M.~{Halpern}, P.~C.
  {Hargrave}, D.~H. {Hughes}, J.~{Klein}, P.~{Mauskopf}, B.~{Magnelli},
  L.~{Moncelsi}, C.~B. {Netterfield}, H.~{Ngo}, L.~{Olmi}, E.~{Pascale},
  G.~{Patanchon}, M.~{Rex}, D.~{Scott}, C.~{Semisch}, N.~{Thomas}, M.~D.~P.
  {Truch}, C.~{Tucker}, G.~S. {Tucker}, M.~P. {Viero}, and D.~V. {Wiebe}, {\it
  {BLAST: Resolving the Cosmic Submillimeter Background}},  {\em \apj} {\bf
  707} (Dec., 2009) 1729--1739, [\href{http://xxx.lanl.gov/abs/0904.1205}{{\tt
  arXiv:0904.1205}}].

\bibitem{Zemcov2010}
M.~{Zemcov}, A.~{Blain}, M.~{Halpern}, and L.~{Levenson}, {\it {Contribution of
  Lensed SCUBA Galaxies to the Cosmic Infrared Background}},  {\em \apj} {\bf
  721} (Sept., 2010) 424--430, [\href{http://xxx.lanl.gov/abs/1006.1360}{{\tt
  arXiv:1006.1360}}].

\bibitem{Renault2001}
C.~{Renault}, A.~{Barrau}, G.~{Lagache}, and J.~{Puget}, {\it {New constraints
  on the cosmic mid-infrared background using TeV gamma-ray astronomy}},  {\em
  \aap} {\bf 371} (May, 2001) 771--778,
  [\href{http://xxx.lanl.gov/abs/astro-ph/}{{\tt astro-ph/}}].

\bibitem{Mazin2007}
D.~{Mazin} and M.~{Raue}, {\it {New limits on the density of the extragalactic
  background light in the optical to the far infrared from the spectra of all
  known TeV blazars}},  {\em \aap} {\bf 471} (Aug., 2007) 439--452,
  [\href{http://xxx.lanl.gov/abs/astro-ph/}{{\tt astro-ph/}}].

\bibitem{Elbaz1999}
D.~{Elbaz}, C.~J. {Cesarsky}, D.~{Fadda}, H.~{Aussel}, F.~X. {D{\'e}sert},
  A.~{Franceschini}, H.~{Flores}, M.~{Harwit}, J.~L. {Puget}, J.~L. {Starck},
  D.~L. {Clements}, L.~{Danese}, D.~C. {Koo}, and R.~{Mandolesi}, {\it {Source
  counts from the 15 mu m ISOCAM Deep Surveys}},  {\em \aap} {\bf 351} (Nov.,
  1999) L37--L40, [\href{http://xxx.lanl.gov/abs/astro-ph/}{{\tt astro-ph/}}].

\bibitem{Pearson2010}
C.~P. {Pearson}, S.~{Oyabu}, T.~{Wada}, H.~{Matsuhara}, H.~M. {Lee}, S.~J.
  {Kim}, T.~{Takagi}, T.~{Goto}, M.~S. {Im}, S.~{Serjeant}, M.~G. {Lee}, J.~W.
  {Ko}, G.~J. {White}, and O.~{Ohyama}, {\it {Source counts at 15 microns from
  the AKARI NEP survey}},  {\em \aap} {\bf 514} (May, 2010) A8+,
  [\href{http://xxx.lanl.gov/abs/1003.2904}{{\tt arXiv:1003.2904}}].

\bibitem{Dole2003}
H.~{Dole}, G.~{Lagache}, and J.~{Puget}, {\it {Predictions for Cosmological
  Infrared Surveys from Space with the Multiband Imaging Photometer for
  SIRTF}},  {\em \apj} {\bf 585} (Mar., 2003) 617--629,
  [\href{http://xxx.lanl.gov/abs/astro-ph/}{{\tt astro-ph/}}].

\bibitem{Dole2004}
H.~{Dole}, G.~H. {Rieke}, G.~{Lagache}, J.~{Puget}, A.~{Alonso-Herrero},
  L.~{Bai}, M.~{Blaylock}, E.~{Egami}, C.~W. {Engelbracht}, K.~D. {Gordon},
  D.~C. {Hines}, D.~M. {Kelly}, E.~{Le Floc'h}, K.~A. {Misselt}, J.~E.
  {Morrison}, J.~{Muzerolle}, C.~{Papovich}, P.~G. {P{\'e}rez-Gonz{\'a}lez},
  M.~J. {Rieke}, J.~R. {Rigby}, G.~{Neugebauer}, J.~A. {Stansberry}, K.~Y.~L.
  {Su}, E.~T. {Young}, C.~A. {Beichman}, and P.~L. {Richards}, {\it {Confusion
  of Extragalactic Sources in the Mid- and Far-Infrared: Spitzer and Beyond}},
  {\em \apjs} {\bf 154} (Sept., 2004) 93--96.

\bibitem{Altieri2010}
B.~{Altieri}, S.~{Berta}, D.~{Lutz}, J.~{Kneib}, L.~{Metcalfe}, P.~{Andreani},
  H.~{Aussel}, A.~{Bongiovanni}, A.~{Cava}, J.~{Cepa}, L.~{Ciesla},
  A.~{Cimatti}, E.~{Daddi}, H.~{Dominguez}, D.~{Elbaz}, N.~M. {F{\"o}rster
  Schreiber}, R.~{Genzel}, C.~{Gruppioni}, B.~{Magnelli}, G.~{Magdis},
  R.~{Maiolino}, R.~{Nordon}, A.~M. {P{\'e}rez Garc{\'{\i}}a}, A.~{Poglitsch},
  P.~{Popesso}, F.~{Pozzi}, J.~{Richard}, L.~{Riguccini}, G.~{Rodighiero},
  A.~{Saintonge}, P.~{Santini}, M.~{Sanchez-Portal}, L.~{Shao}, E.~{Sturm},
  L.~J. {Tacconi}, I.~{Valtchanov}, M.~{Wetzstein}, and E.~{Wieprecht}, {\it
  {Herschel deep far-infrared counts through Abell 2218 cluster-lens}},  {\em
  \aap} {\bf 518} (July, 2010) L17+,
  [\href{http://xxx.lanl.gov/abs/1005.1575}{{\tt arXiv:1005.1575}}].

\bibitem{Condon1974}
J.~J. {Condon}, {\it {Confusion and Flux-Density Error Distributions}},  {\em
  \apj} {\bf 188} (Mar., 1974) 279--286.

\bibitem{Friedmann2004}
Y.~{Friedmann} and F.~{Bouchet}, {\it {Fluctuation analysis of the far-infrared
  background - information from the confusion}},  {\em \mnras} {\bf 348} (Mar.,
  2004) 737--744, [\href{http://xxx.lanl.gov/abs/astro-ph/}{{\tt astro-ph/}}].

\bibitem{Patanchon2009}
G.~{Patanchon}, P.~A.~R. {Ade}, J.~J. {Bock}, E.~L. {Chapin}, M.~J. {Devlin},
  S.~R. {Dicker}, M.~{Griffin}, J.~O. {Gundersen}, M.~{Halpern}, P.~C.
  {Hargrave}, D.~H. {Hughes}, J.~{Klein}, G.~{Marsden}, P.~{Mauskopf},
  L.~{Moncelsi}, C.~B. {Netterfield}, L.~{Olmi}, E.~{Pascale}, M.~{Rex},
  D.~{Scott}, C.~{Semisch}, N.~{Thomas}, M.~D.~P. {Truch}, C.~{Tucker}, G.~S.
  {Tucker}, M.~P. {Viero}, and D.~V. {Wiebe}, {\it {Submillimeter Number Counts
  from Statistical Analysis of BLAST Maps}},  {\em \apj} {\bf 707} (Dec., 2009)
  1750--1765, [\href{http://xxx.lanl.gov/abs/0906.0981}{{\tt
  arXiv:0906.0981}}].

\bibitem{Glenn2010}
J.~{Glenn}, A.~{Conley}, M.~{B{\'e}thermin}, B.~{Altieri}, A.~{Amblard},
  V.~{Arumugam}, H.~{Aussel}, T.~{Babbedge}, A.~{Blain}, J.~{Bock},
  A.~{Boselli}, V.~{Buat}, N.~{Castro-Rodr{\'{\i}}guez}, A.~{Cava},
  P.~{Chanial}, D.~L. {Clements}, L.~{Conversi}, A.~{Cooray}, C.~D. {Dowell},
  E.~{Dwek}, S.~{Eales}, D.~{Elbaz}, T.~P. {Ellsworth-Bowers}, M.~{Fox},
  A.~{Franceschini}, W.~{Gear}, M.~{Griffin}, M.~{Halpern},
  E.~{Hatziminaoglou}, E.~{Ibar}, K.~{Isaak}, R.~J. {Ivison}, G.~{Lagache},
  G.~{Laurent}, L.~{Levenson}, N.~{Lu}, S.~{Madden}, B.~{Maffei},
  G.~{Mainetti}, L.~{Marchetti}, G.~{Marsden}, H.~T. {Nguyen}, B.~{O'Halloran},
  S.~J. {Oliver}, A.~{Omont}, M.~J. {Page}, P.~{Panuzzo}, A.~{Papageorgiou},
  C.~P. {Pearson}, I.~{P{\'e}rez-Fournon}, M.~{Pohlen}, D.~{Rigopoulou},
  D.~{Rizzo}, I.~G. {Roseboom}, M.~{Rowan-Robinson}, M.~S. {Portal},
  B.~{Schulz}, D.~{Scott}, N.~{Seymour}, D.~L. {Shupe}, A.~J. {Smith}, J.~A.
  {Stevens}, M.~{Symeonidis}, M.~{Trichas}, K.~E. {Tugwell}, M.~{Vaccari},
  I.~{Valtchanov}, J.~D. {Vieira}, L.~{Vigroux}, L.~{Wang}, R.~{Ward},
  G.~{Wright}, C.~K. {Xu}, and M.~{Zemcov}, {\it {HerMES: deep galaxy number
  counts from a P(D) fluctuation analysis of SPIRE Science Demonstration Phase
  observations}},  {\em \mnras} {\bf 409} (Nov., 2010) 109--121,
  [\href{http://xxx.lanl.gov/abs/1009.5675}{{\tt arXiv:1009.5675}}].

\bibitem{Fazio1970}
G.~G. {Fazio} and F.~W. {Stecker}, {\it {Predicted High Energy Break in the
  Isotropic Gamma Ray Spectrum: a Test of Cosmological Origin}},  {\em \nat}
  {\bf 226} (Apr., 1970) 135--136.

\bibitem{Raue2009}
M.~{Raue}, T.~{Kneiske}, and D.~{Mazin}, {\it {First stars and the
  extragalactic background light: how recent {$\gamma$}-ray observations
  constrain the early universe}},  {\em \aap} {\bf 498} (Apr., 2009) 25--35,
  [\href{http://xxx.lanl.gov/abs/0806.2574}{{\tt arXiv:0806.2574}}].

\bibitem{Kneiske2010}
T.~M. {Kneiske} and H.~{Dole}, {\it {A lower-limit flux for the extragalactic
  background light}},  {\em \aap} {\bf 515} (June, 2010) A19+,
  [\href{http://xxx.lanl.gov/abs/1001.2132}{{\tt arXiv:1001.2132}}].

\bibitem{Planck2011} Planck 
Collaboration, et al.\ 2011, arXiv:1101.2028 

\bibitem{Amblard2011} Amblard, A., et al.\ 
2011, arXiv:1101.1080

\end{thebibliography}\endgroup


\providecommand{\href}[2]{#2}\begingroup\raggedright\endgroup

\end{document}